\documentclass[12pt, preprint]{aastex}

\begin{document}

\title{Shocked Molecular Hydrogen in the 3C 326 Radio Galaxy System}

\author{Patrick Ogle$^1$, Robert Antonucci$^2$, P. N. Appleton$^3$, \& David Whysong$^4$}

\affil{$^1$Spitzer Science Center, California Institute of Technology, 
       Mail Code 220-6, Pasadena, CA 91125}

\affil{$^2$Physics Dept., University of California, Santa Barbara, CA 93106}

\affil{$^3$NASA Herschel Science Center, California Institute of Technology, 
       Mail Code 100-22, Pasadena, CA 91125}

\affil{$^4$ NRAO, Array Operations Center, P. O. Box O, 1003 Lopezville Rd., Socorro
, NM 87801-0387}

\email{ogle@ipac.caltech.edu}

\shorttitle{Shocked Molecular Hydrogen in 3C 326}
\shortauthors{Ogle et al.}

\begin{abstract}

The {\it Spitzer} spectrum of the giant FR II radio galaxy 3C 326 
is dominated by very strong molecular hydrogen emission lines on a faint IR continuum. 
The H$_2$ emission originates in the northern component of a double-galaxy system associated 
with 3C 326. The integrated luminosity in H$_2$ pure-rotational lines is $8.0\times 10^{41}$ 
erg s$^{-1}$, which corresponds to 17\%  of the 8-70 $\mu$m luminosity of the galaxy. A wide 
range of temperatures (125-1000 K) is measured from the H$_2$ 0-0 S(0)-S(7) transitions,
leading to a {\it warm} H$_2$ mass of $1.1\times 10^9 M_\odot$. Low-excitation ionic forbidden 
emission lines are consistent with an optical LINER classification for the active nucleus, which is 
{\it not} luminous enough to power the observed H$_2$ emission. The H$_2$ could be 
shock-heated by the radio jets, but there is no direct indication of this. More likely, the H$_2$ 
is shock-heated in a tidal accretion flow induced by interaction with the southern companion 
galaxy. The latter scenario is supported by an irregular morphology, tidal bridge, and possible 
tidal tail imaged with IRAC at 3-9 $\mu$m. Unlike ULIRGs, which in some cases exhibit H$_2$ line 
luminosities of comparable strength, 3C 326 shows little star-formation activity 
($\sim 0.1 ~M_\odot$ yr$^{-1}$). This may represent an important stage in galaxy evolution. 
Starburst activity and efficient accretion onto the central supermassive black hole may be delayed 
until the shock-heated H$_2$ can kinematically settle and cool.

\end{abstract}


\section{Introduction}

Recent {\it Spitzer} and ISO observations have revealed a new class of extremely luminous molecular 
hydrogen emission galaxies (MOHEG's) with $L$(H$_2)= 10^{40}-10^{43}$ erg s$^{-1}$ in pure-rotational 
molecular hydrogen emission lines but relatively weak total IR emission, $L$(H$_2$)/$L$(IR)$>10^{-3}$. 
In comparison, normal star-forming galaxies in the SINGS survey of the local universe have 
$L$(H$_2$)/$L$(IR)$\sim 6 \times 10^{-4}$ \citep{rhh07}.

\cite{a06} discovered H$_2$ emission extended over $50\arcsec$ in the giant intergalactic X-ray shock
front in the Stephan's Quintet compact group. The estimated H$_2$ luminosity from the whole shock 
is in excess of 10$^{41}$ ergs/s. Nearly 30\% of the IR luminosity is in H$_2$ lines, while the rest is 
emitted by $\sim 40$ K dust in the far-IR \citep{xu03}. The power in the H$_2$ lines compared with the
dust continuum is several orders of magnitude larger than that seen in photo-dissociation regions (PDRs).
The H$_2$ emission lines are very broad (870 km s$^{-1}$), suggesting turbulent motions inside the shock 
front. The H$_2$ emission may be powered by non-dissociative shocks in dense clumps or filaments inside 
the shock front. 

\cite{e06} find extremely luminous $L$(H$_2$)$= 2\times 10^{43}$ erg s$^{-1}$ emission from 
$\sim 10^{10} M_\odot$ of warm H$_2$ in the brightest galaxy of the high-luminosity X-ray cluster Zw 3146. 
This is accompanied by high-luminosity dust emission powered by starburst activity. Lower luminosity 
ro-vibrational H$_2$ emission is similarly seen in several low-redshift galaxy clusters 
\citep{d00,hcj05,jhf07}. The source of excited H$_2$ in galaxy clusters is an unsolved mystery--cooling 
flows, jet shocks, and X-ray heating by active galactic nuclei (AGNs) have all been considered. 

Molecular hydrogen emission is often seen in interacting galaxies and ultra-luminous infrared galaxies (ULIRGs), but 
$L$(H$_2$)/$L$(IR) is typically $<<10^{-3}$ \citep{hck05, hah06}. The late-stage merger/LIRG NGC 6240 
has atypically luminous  emission from $1.6\times 10^9 M_\odot$ of warm H$_2$, accompanied by a prodigious 
star-formation rate of $\sim 120 ~M_\odot$ yr$^{-1}$ \citep{lsg03,nlk05,mcm05,abs06}. The overlap region between the 
Antennae colliding galaxies also shows unusually strong H$_2$ emission with $L$(H$_2$)/$L$(IR)$\sim 1\times 10^{-3}$, 
possibly indicating a pre-starburst shock \citep{hck05}.

We present the remarkable H$_2$ emission spectrum of the FR II radio galaxy 3C 326, observed with the {\it Spitzer} 
Space Telescope. This is part of a large survey of 52 FR II radio galaxies from the 3CRR catalog \citep{owa06},
and another survey of 21 FR I radio galaxies \citep{oaw07}. So far, we have found 3 FR IIs (6\%) and 5 FR Is (24\%) 
with strong H$_2$ emission (Ogle et al. 2007, in preparation). Of the galaxies in our sample, 3C 326 has the largest 
H$_2$ emission line equivalent widths. The weak mid-IR continuum of 3C 326 indicates the presence of a radiatively 
inefficient AGN, in spite of its large radio power and enormous size of its radio lobes. 

\section{Radio Galaxy Identification}

The radio source 3C 326 is one of the largest known \citep{ws78}, with an angular size of 19\farcm5 (1.9 Mpc; we assume 
a cosmology with $H_0=70$ km s$^{-1}$ Mpc$^{-1}$, $\Omega_\mathrm{m}=0.3$, and $\Omega_\Lambda=0.7$). The pair of 
galaxies 3C 326 N and 3C 326 S are both at a redshift of $z=0.089\pm0.001$ and have a projected separation of 
24\farcs9 (41 kpc). There is some confusion concerning which of the two galaxies hosts the AGN that produced 
the Mpc-scale radio lobes. {\it Both} galaxies have unresolved ($<3\arcsec$) radio sources coincident with their 
nuclei, with $S_\nu(8 ~\mathrm{GHz})=1.5$ and 6.5 mJy, respectively \citep{rsm90}. While the southern radio source 
is brighter, both sources are consistent with typical FR II radio galaxy cores. The galaxy 3C 326 N has a 
LINER-like optical emission line spectrum (strong [O{\sc ii}], H$\alpha$, [S {\sc ii}], and weak [O{\sc iii}]), 
while 3C 326 S has no detected emission lines \citep{rsm90,swc96}. Also, the K-band magnitude and corresponding 
stellar mass of 3C 326 N is more typical of FR II radio galaxy hosts \citep{rsm90}. We therefore targeted 3C 326 N 
with the {\it Spitzer} IRS. 

\section{Observations}

\subsection{Spitzer IRS}

The radio galaxy 3C 326 N was observed with the Spitzer IRS \citep{h04}, covering the wavelength range 
5-35 $\mu$m. The companion galaxy 3C 326 S was observed on one nod of the Short-Low (SL) slits because of a fortuitous 
slit PA (Fig.1). The slit widths in the low-resolution modules Short-Low 2 (SL2),  Short-Low 1 (SL1), Long-Low 2 
(LL2), and Long-Low 1 (LL1) are 3\farcs6, 3\farcs7, 10\farcs5, and 10\farcs7, respectively.  The SL and LL spectral 
resolutions range from 60-130. The exposure times were 240 seconds in each spectral order except LL2 (480 seconds).

Our data reductions began from the S15.3.0 pipeline-processed basic calibrated data (BCDs). Off-slit observations were 
subtracted to remove background light. The 2D-spectra were cleaned using IRSCLEAN 1.8 to median filter bad pixels. 
Spectra were extracted using SPICE 1.4 within tapered regions matching the {\it Spitzer} point-spread function  
(SL2: 7\farcs2 at 6 $\mu$m, SL1: 14\farcs4 at 12 $\mu$m, LL2: 21\farcs7 at 16 $\mu$m, LL1: 36\farcs6 at 27 $\mu$m). 
The second LL2 nod was considerably noisier than the first, so we discarded this data. Optimal extraction (which
assumes a point-source spatial profile) yielded significant gains in S/N  for both galaxies. 

At the redshift of 3C 326, 1\farcs0 corresponds to 1.7 kpc. The SL and LL slit widths correspond to 6.2 kpc and 
18 kpc, respectively, and are considerably smaller than the optical extents of the two galaxies. However, the 
continuum level is well-matched across orders and the emission lines and dust features do not appear to be spatially 
extended along the slit for either galaxy (Figs. 1, 2, 3). The continuum and emission line fluxes from the optimally 
extracted spectra also agree well with the regular extractions, indicating that the galaxies are (excepting the
stellar component at short wavelengths) mostly unresolved by {\it Spitzer} IRS, and that aperture effects are unimportant
for studying the dust and gas emission. 

\subsection{The Mid-IR Spectra}

The spectrum of 3C 326 N is dominated by pure-rotational emission lines from molecular hydrogen (Fig. 2, Table 1).
We detect the full series of emission lines from H$_2$ 0-0 S(0)-S(7). Forbidden emission lines from the low-moderate 
ionization species Fe {\sc ii}, S {\sc iv}, Ne {\sc ii}, S {\sc iii}, and O {\sc iv} are also observed (Table 2).
Polycyclic aromatic hydrocarbon (PAH) emission features are seen at 6.7, 7.7, 10.7, 11.3, 13.6, and 17 $\mu$m (Table 3). 
We consider the PAHs at 6.7, 7.7, 10.7, 15.9 and 18.9 $\mu$m and the Fe {\sc ii}, S {\sc iv}, and S {\sc iii} lines to
be marginal ($<3\sigma$) detections. 

We model the 3C 326 N spectrum using the $\chi^2$ fitting routine PAHFIT \citep{sdd07}.
Extinction is not included in the model since it does not improve the fit. Emission lines are fit by Gaussians and 
PAH features are fit by Drude profiles (Tables 1-3). None of the emission lines are resolved, all with 
FWHM$<2300-5000$ km s$^{-1}$. The 4.8-8 $\mu$m continuum is dominated by galaxian starlight. 
An excess over the stellar component in the 8-24 $\mu$m region is fit by thermal emission from 
warm (90-300 K) dust, with integrated flux $F_{8-24}=9.0\pm 0.9 \times 10^{-14}$ erg s$^{-1}$ cm$^{-2}$
and luminosity $L_{8-24}=1.8\pm 0.2 \times 10^{42}$ erg s$^{-1}$.

The companion galaxy 3C 326 S has no H$_2$ emission detected; the 3$\sigma$ upper limit to the H$_2$ S(3) line
flux is $1.2\times 10^{-15}$ erg s$^{-1}$ cm$^{-2}$ ($L$(H$_2$)$<2.4 \times 10^{40}$ erg s$^{-1}$). No ionic forbidden 
emission lines are detected, consistent with the lack of optical emission lines and indicating an inactive nucleus. 
A spike in the spectrum at $\sim 12$ $\mu$m does not match any known emission lines and may result from a hot pixel. 
The 7.7 and 11.3 PAH features are 2-3 times as strong in 3C 326 S compared to 3C 326 N (Table 3) and the mean 9-10 
$\mu$m continuum flux density ($0.62 \pm 0.04$ mJy) is 20\% brighter.

\subsection{Spitzer IRAC and MIPS}

We retrieved the IRAC \citep{f04} 3.6, 4.5, 5.8, and 8.0 $\mu$m S14.0.0-processed images from the {\it Spitzer} archive
(Program-ID 03418, PI M. Birkinshaw) to further constrain the spatial distribution of the stellar, dust, and molecular gas 
components (Figs. 3,4). A 3$\sigma$ clipped average of the dither positions, excluding frames affected by a bright star at 
the edge of the field, was used to subtract the background light, and frames were combined using the SSC MOPEX-MOSAIC 
software. The 3.6 and 4.5 $\mu$m bands measure the stellar continuum. The 5.8 $\mu$m band contains the redshifted H$_2$ 
S(5)-S(7) lines and 6.7 $\mu$m PAH feature. The 8.0 $\mu$m band contains the redshifted H$_2$ S(3)-S(4) lines and 7.7 $\mu$m 
PAH feature. Both of the latter two bands also contain thermal continuum emission from warm dust. 

The irregular morphology of 3C 326 N in the IRAC images suggests an ongoing interaction or merger with 3C 326 S 
($24\farcs9$ at PA$=177\arcdeg$), a relatively compact elliptical galaxy. The isophotes of the two galaxies merge and 
they appear to be connected by a tidal bridge. There is a possible tidal tail extending from the NW quadrant at 
PA$=290-334\arcdeg$. Two other sources, one $4\farcs3$ away from the center of 3C 326 at PA$=111\arcdeg$ and one 
$15\farcs9$ away at PA$=268\arcdeg$, appear to be unrelated background galaxies seen in projection, based on their 
morphologies, red colors, and small angular diameters in archival Sloan Digital Sky Survey (SDSS) images 
\footnote{http://www.sdss.org/}.

We also obtained 24, 70, and 160 $\mu$m MIPS \citep{r04} imaging data from the Spitzer archive for 3C 326 (Fig. 5). 
Individual BCD frames from the S14.4 pipeline were processed with the MOPEX-MOSAIC software to generate final images. 
Aperture photometry was performed on both 3C 326 N and 3C 326 S, and point-source aperture corrections\footnote{obtained 
from http://ssc.spitzer.caltech.edu} were applied to derive the fluxes (Table 4 and Fig. 6). Both galaxies are detected at 
24 and 70, but not 160 $\mu$m. We measure a root-mean-square noise of 0.4 MJy sr$^{-1}$ and $3\sigma$ flux upper limits 
of 34 mJy at 160 $\mu$m. This is consistent with a non-detection at 870 $\mu$m \citep{qay03}, within
a $23 \arcsec$ beam centered on 3C 326 S.

The companion 3C 326 S is twice as bright as 3C 326 N at 24 $\mu$m, while 3C 326 N is 40\% brighter at 70 $\mu$m.
A third source is seen in the 70 $\mu$m image at the location where the tidal bridge in the IRAC 3.6 $\mu$m image 
connects to 3C 326 N (Fig.5, Table 4). It is not detected in the MIPS 24 $\mu$m band, nor in any of the IRAC bands. If it is 
truly associated with 3C 326, then it may indicate cool dust in the tidal bridge. Unfortunately, none of the IRS slits covered
this feature, so we do not know if there is any associated H$_2$ emission.

The spectral energy distributions (SEDs) of 3C 326 N and S are similar, but do show interesting differences (Fig. 6). Both are
dominated by stellar emission at frequencies $>4\times 10^{13}$ Hz (wavelengths $<8$ $\mu$m), consistent with our model of the
IRS spectrum of 3C 326 N. Dust emission contributes most of the flux at $\sim 1\times 10^{12}-4\times 10^{13}$ Hz and synchrotron 
emission from the radio cores is seen at 3-10 GHz. The ratio $L$(70 $\mu$m)/$L$(24 $\mu$m) is greater in 3C 326 N than in 3C 
326 S (Table 4), indicating relatively more cool dust emission and relatively less warm dust emission. 

We estimate far-IR continuum luminosities by interpolating the 24 and 70 $\mu$m photometric points with a power law. 
This yields $L_{24-70}=3.0\pm 0.4 \times 10^{42}$ erg s$^{-1}$ and $3.3\pm 1.8 \times 10^{42}$ erg s$^{-1}$, respectively, 
for 3C 326 N and 3C 326 S. Adding together the MIR 8-24 $\mu$m and FIR 24-70 $\mu$m emission, the 8-70 $\mu$m luminosity of 
3C 326 N is $L_{8-70}=4.8 \pm 0.5 \times 10^{42}$ erg s$^{-1}$ ($1.2\pm 0.1 \times 10^9 L_\odot$). Upper and lower bounds to the 
FIR luminosity at longer wavelengths of $L_{70-1000}=0.3-1.7\times 10^{43}$ erg s$^{-1}$ are estimated for either galaxy by 
assuming blackbody emission that peaks at either 70 $\mu$m or 160 $\mu$m (Fig. 6). The total 8-1000 $\mu$m IR luminosity of 
3C 326 N therefore lies in the range $L_\mathrm{IR}=0.8-2.2 \times 10^{43}$ erg s$^{-1}$ ($2-6\times 10^9 L_\odot$).

\section{H$_2$ Emission}

The galaxy 3C 326 N has H$_2$ pure-rotational lines of extraordinary luminosity and equivalent width (Fig. 2, Table 1).
The H$_2$ S(3) line has a luminosity of $2.34\pm 0.08 \times 10^{41}$ erg s$^{-1}$ and an equivalent width 
of $0.80 \pm 0.08$ $\mu$m. The integrated S(0)-S(7) line luminosity is  $L(\mathrm{H}_2)=8.0\pm 0.4 \times 10^{41}$ erg s$^{-1}$.
This is $17\pm 2\%$ of the measured 8-70 $\mu$m luminosity and $4-10\%$ of the estimated 8-1000 $\mu$m luminosity, making it the 
most extreme H$_2$-emitting galaxy seen so far by Spitzer (Table 6). Unlike the giant, extended, intergalactic shock in 
Stephan's Quintet, the H$_2$ emission here is unresolved and contained within the central region of 3C 326 N. 

We estimate the upper-level column densities ($N/g$) for each observed H$_2$ transition, divided by the statistical 
weights (Fig. 7, Table 1), assuming that the H$_2$ source just fits inside the SL slit (3\farcs7).  The column density 
distribution is fit by a minimum of 3 temperature components with $T=125,400,$ and 1000 K (Table 5). The fit 
is not unique but rather serves to demonstrate the large range of temperatures of the H$_2$ emission regions. The
total column density of 125 K H$_2$ averaged over the SL slit is $1.8\times 10^{21}$ cm$^{-2}$.  

The ratio of ortho (odd angular momentum quantum number J) to para (even J) H$_2$ depends on the gas temperature 
and thermal history. Warm H$_2$ in thermal equilibrium at a temperature of 125 K should have ortho/para=2.1, which is 
consistent with our model fit. The ratio of ortho/para H$_2$ in the hot (400 K and 1000 K) components is consistent with
the standard value of 3.0 for gas in thermal equilibrium at $T>250$ K \citep{wcp00,nms06}.

The masses of {\it warm} and {\it hot} molecular gas sum to $1.1\times 10^9 M_\odot$ (Table 6). This is 
comparable to the warm H$_2$ mass in NGC 6240 \citep{lsg03,abs06} and to the {\it total} molecular 
gas mass of the Milky Way ($\sim 2 \times 10^9 M_\odot$).  No published CO observations of 3C 326 are
available, so we have no knowledge of the mass of cold H$_2$; although by analogy with galaxies like NGC 6240
the cold component could be even larger than the warm and hot components. In thermodynamic equilibrium, 
the H$_2$ component densities must exceed the critical densities of the observed rotational transitions \citep{lpf99}.  
This yields maximum filling factors ranging from $f_\mathrm{max}=2\times 10^{-3}$ for the 125 K component to 
$f_\mathrm{max}=1\times 10^{-9}$ for the 1000 K component within the central 6.2 kpc-diameter sphere (Table 5). 

\section{Forbidden Emission Lines and AGN Activity}

The mid-IR forbidden emission lines from several ions (Table 2) can be used to distinguish between starburst and
AGN activity. In particular, line flux ratios of  [O {\sc iv}]/[Ne {\sc ii}] $=0.6\pm 0.1$ and
[S {\sc iv}]/[Ne {\sc ii}] $=0.08\pm 0.04$ in 3C 326 are characteristic of a LINER AGN \citep{sm92,src06}. 
Seyfert galaxies have relatively stronger [O {\sc iv}] and [S {\sc iv}] than this, while starburst galaxies are
relatively weaker in these two lines. Consequently, little of the forbidden line emission comes from 
star-formation activity. A LINER classification for 3C 326 is also consistent with the optical line ratio 
[O {\sc iii}] 5007\AA/[O {\sc ii}] 3727\AA $=0.30$ \citep{swc96}.

{\it Spitzer} observations give an {\it upper limit} to the IR luminosity of the AGN, since there is a significant
contribution from the host galaxy.  The fiducial 15 $\mu$m continuum flux is $0.87 \pm 0.09$ mJy, 
somewhat greater than the upper limit measured by \cite{owa06}. (Our measurement accuracy has improved by throwing out 
the noisy LL2 nod and using optimal extraction.) The corresponding luminosity is 
$\nu L(15 ~\mu\mathrm{m})=3.4 \pm 0.4 \times 10^{42}$ erg s$^{-1}$ and its ratio to the radio lobe luminosity is only 
$\nu L(15 ~\mu$m)/$\nu L$(178 MHz)$=4.4$. We confirm that this powerful radio galaxy contains an AGN with relatively 
low accretion power. In comparison, we find from archival {\it Spitzer} IRS observations that the LINER/FR I radio 
galaxy M 87 is one tenth as bright, with  $\nu L(15 ~\mu\mathrm{m})=3.3 \pm 0.1 \times 10^{41}$ erg s$^{-1}$, and  has
a similar ratio of $\nu L(15 \mu\mathrm{m})/\nu L(178 \mathrm{MHz})=3.9$. It has been argued that low-power radio
galaxies such as M 87 are fueled by radiatively inefficient accretion \citep{rdf96}.

X-ray heating has been proposed as a possible mechanism to produce H$_2$ emission in active galaxies \citep{kl89,dw90,rkl02}. 
Unfortunately, there are no published X-ray observations of 3C 326 N. However, \cite{hec06} find a $\sim$1:1 correlation 
between 15 $\mu$m continuum emission and accretion-powered 2-10 keV X-ray luminosity for $z<0.5$ radio galaxies. If this 
correlation holds true for 3C 326 N and the H$_2$ were powered by X-rays, then the observed ratio of 
$L$(H$_2$)/$L$(15 $\mu$m)$=0.23\pm 0.03$ would require an implausibly high conversion of X-rays to H$_2$ emission. 
{\it Chandra} X-ray observations are necessary to directly confirm this, and will also be useful for measuring the 
high-energy SED of the AGN to test accretion models.

\section{Aromatic Feature Emission and Star Formation}

We detect several aromatic (PAH) emission features in the IRS spectrum of 3C 326 N (Fig.2, Table 3).
The  7.7 and 11.3 $\mu$m PAH luminosities are $1.0 \pm 0.5 \times 10^{41}$ erg s$^{-1}$ and 
$1.3 \pm 0.3 \times 10^{41}$ erg s$^{-1}$, respectively. The 11.3/7.7 PAH flux ratio is $1.2\pm 0.6$,
to be compared with an average ratio of $0.3\pm0.1$ for spiral galaxies and AGN host galaxies 
\citep{sdd07,sor07}.  The total luminosity in detected PAH features is  $L(\mathrm{PAH})=1.0 \pm 0.1 \times 10^{42}$ 
erg s$^{-1}$ and the ratio $L$(H$_2$)/$L$(PAH)$=0.8\pm 0.1$. This ratio is much greater than the typical
ratio of  $\sim 1\times 10^{-2}$ seen in normal star-forming galaxies where both H$_2$ and PAHs are thought to 
be produced in stellar PDRs \citep{rhh07}.

PAH emission excited by UV photons from O and B stars roughly correlates with other star-formation rate (SFR) indicators 
such as H$\alpha$ luminosity and 24 $\mu$m continuum luminosity \citep{rsv01}. However, there is a large scatter, with 
PAH luminosity depending on metallicity and age of the stellar population \citep{cke07}. X-ray 
radiation from an AGN can destroy PAHs \citep{v91}, but should have little effect at large distances or in regions 
shielded from the nucleus. With these caveats in mind, we estimate the SFR from the luminosity of the 7.7 $\mu$m PAH 
feature to be $7 \pm 3 \times 10^{-2} ~M_\odot$ yr$^{-1}$ for 3C 326 N. The 24 $\mu$m luminosity measured with MIPS 
yields a comparable SFR of $0.14 \pm 0.02 ~M_\odot$ yr$^{-1}$. The similarity of these two estimates indicates that the 
AGN does not dominate the continuum at this wavelength ($\lesssim 50\%$). The star formation rate is only $2-5\%$ of 
the $\sim 3 ~M_\odot$ yr$^{-1}$ SFR of the Milky Way.

A remarkable aspect of 3C 326 N (and MOHEGs in general) is the lack of starburst activity in the presence of a large 
molecular gas mass. The mean surface density of warm H$_2$ inside the SL slit is $30 ~M_\odot$ pc$^{-2}$. In order to 
suppress star formation in a central thin disk, the total gas surface density must be below the critical density 
\citep{k89}. An accurate estimate of the critical density in 3C 326 N will require a kinematic study of the 
molecular gas. A large turbulent velocity dispersion or rotational velocity shear would increase the critical density 
for star formation. The molecular gas may have not yet even settled into a disk. Efficient star 
formation and AGN accretion may be suppressed and delayed in 3C 326 N until the molecular gas has kinematically settled 
and cooled. 

The PAH and 24 $\mu$m luminosities of companion galaxy 3C 326 S yield estimated SFRs of $0.20\pm 0.05 ~M_\odot$ 
yr$^{-1}$ and $0.26\pm 0.03 ~M_\odot$ yr$^{-1}$, respectively. These SFR estimates are within the broad range seen for 
normal, non-starburst early type galaxies \citep{cyb07}. The SFR in 3C 326 S is significantly larger than the SFR in
3C 326 N. Since no H$_2$ emission lines are detected, it is likely that most of the remaining molecular gas is cold and 
conducive to star formation. A {\it Hubble} Space Telescope image of 3C 326 S in the F702W band shows a dust lane (or nearly 
edge-on disk) cutting across the core at PA$\sim 105\arcdeg$, which confirms the presence of a significant dusty ISM 
\citep{mbs99}. The galaxy 3C 326 S may be the H$_2$ donor for 3C 326 N, and may have had a much larger molecular gas component 
which has been largely stripped by tidal interaction. It would be particularly useful to measure the masses and surface 
densities of cold CO (and thereby H$_2$) in both galaxies to see if they are consistent with the observed SFRs and to help 
understand why the warm H$_2$ emission is so much stronger in 3C 326 N.

\section{Energetics and H$_2$ Heating Mechanisms}

In 3C 326 N, the ratio of mid-IR rotational H$_2$ emission to the 8-70 $\mu$m IR continuum emission is 
$L$(H$_2$)/$L_{8-70}=0.17 \pm 0.02$. The H$_2$ pure rotational lines are therefore a major coolant for the warm molecular 
gas phase of the ISM. The 125 K H$_2$ component has a thermal energy content of $\sim 4\times 10^{52}$ erg and the cooling 
time from the observed luminosity in H$_2$ S(0)+S(1) lines is only $9 \times 10^{3}$ yr. It is therefore necessary to 
continuously inject energy into the molecular gas to maintain the observed temperature. We consider below two possible 
scenarios for heating the H$_2$, i) shock-heating by the inner radio jet and ii) tidally-induced inflow or accretion 
from the companion galaxy 3C 326 S. 

\subsection{Radio Jet Heating Scenario}

The radio lobe size and Alfv\'en speed give an estimated expansion age of $\sim 2\times 10^8$ yr for 
the Mpc-scale radio source \citep{ws78}. The terminal working surfaces of the radio jets are in intergalactic space, 
far outside both galaxies. Any jet-heating of H$_2$ must be done locally by the inner jet at a radius of $\lesssim 3.1$ kpc. 
The average jet power is best estimated from the extended radio lobe emission. The minimum energy content in the 
radio lobes derived from equipartition arguments is $6.0\times10^{59}$ erg \citep{ws78}. Dividing by the 
radio source lifetime, we estimate a jet kinetic luminosity of  $9.5 \times 10^{43}$ erg s$^{-1}$. We estimate a
smaller kinetic luminosity of  $4.8 \times 10^{43}$ erg s$^{-1}$ if the lobe magnetic fields have sub-equipartition 
strength and the lobe energy is dominated by relativistic particles \citep{p06,p05}. 

The integrated 10 MHz-10 GHz synchrotron luminosity from the radio lobes is $4.7 \times 10^{42}$ erg s$^{-1}$,  $\sim5\%$ 
of the estimated jet kinetic luminosity. The synchrotron luminosity of the 3C 326 N core is much weaker, at 
$3.5 \times 10^{39}$ erg s$^{-1}$. If we assume that the synchrotron luminosity is roughly proportional to the 
dissipation of jet kinetic energy, then the ratio of core to lobe total luminosity indicates that 
$\sim 7\times10^{-4}$ of the jet kinetic luminosity, or $\sim7 \times 10^{40}$ erg s$^{-1}$ is dissipated in 3C 326 N. 
(For the sake of illustration, this assumes that relativistic beaming is not important in the core. In the likely case 
that the core emission is beamed, this should be divided by the beaming factor.) This is less than one tenth
of the luminosity required to power the observed H$_2$ emission. Also, jet shocks must be active over a large fraction 
of the total H$_2$ mass in the galaxy to be a viable heating mechanism. However, with a large uncertainty in the amount 
of jet kinetic power dissipated and the unknown spatial configuration of H$_2$, this scenario may deserve further 
consideration. High-spatial resolution radio observations are necessary to look for any direct evidence of jet-ISM 
interaction.

\subsection{Tidally Induced Inflow}

Another, preferred scenario is that 3C 326 N has been disturbed by gravitational interaction with 3C 326 S. The
H$_2$ may have either been stripped from 3C 326 S, or already present in 3C 326 N. Tidal forces from 3C 326 S would then 
induce the observed distortions in 3C 326 N, in turn giving rise to gravitational torque that drives molecular gas into 
the galaxy center. The gravitational energy of the in-falling gas is then converted into turbulence and heat via an accretion 
shock or shocks between sub-clumps. The simulations of \cite{mh96} show that the greatest gas inflow rates are achieved for 
co-planar interactions of disk systems $\sim 2\times 10^8$ yr after the initial encounter due to the growth of a central bar 
over the disk dynamical time scale, but before the final merger. The tidal field of the bar drives the inflow.

The K-band magnitudes of 3C 326 N and S are 12.95 and 13.60, respectively \citep{ll84, rsm90}. We estimate stellar 
bulge masses of $3.1 \times 10^{11} ~M_\odot$ and $1.6 \times 10^{11} ~M_\odot$, respectively, using an empirical relation between 
K-band luminosity and bulge mass \citep{mh03}. The $\sim$2:1 mass ratio qualifies this as a major interaction. The
corresponding nuclear supermassive black hole masses (assuming $M_\mathrm{BH}/M_\mathrm{bulge}=1.5\times 10^{-3}$) are
$\sim 5 \times 10^{8} ~M_\odot$ and $\sim 2 \times 10^{8} ~M_\odot$, which may eventually result in a major black hole merger
if the two galaxies merge. The isophotes of 3C 326 N are roughly elliptical at a radius of $12\farcs2$, with PA$=154\arcdeg$. At 
larger radii ($12-16\arcsec$), the isophotes bend to the south to meet 3C 326 S, perhaps indicating a tidal bridge. The two 
galaxies have identical radial velocities to within the measurement uncertainties ($\pm 300$ km s$^{-1}$).

The 2-body L1 Lagrange point is at $24 / \cos(\theta)$ kpc from the nucleus of 3C 326 N and $17 / \cos(\theta)$ kpc from 
the nucleus of 3C 326 S, where $\theta$ is the inclination of the orbital axis. The gravitational potential energy released by 
$1.1 \times 10^9 M_\odot$ of H$_2$ falling from L1 to the center of 3C 326 N is $8.4 \times 10^{57}$ erg. Dividing by the 
dynamical (free-fall) time-scale of $t_\mathrm{dyn}=9.1\times 10^7$ yr, we obtain an H$_2$ inflow rate of 12 $M_\odot$ yr $^{-1}$ 
and an accretion luminosity of $2.9\times 10^{42}$ erg s$^{-1}$. This is sufficient to power the observed  H$_2$ emission at a 
conversion efficiency of 27\%. A massive dark halo will deepen the gravitational potential and allow an even lower conversion 
efficiency. The near-coincidence of the crossing time-scale $2 t_\mathrm{dyn} = 1.8\times 10^8$ yr and the age of the radio source 
suggests the intriguing possibility that the radio jet activity is ultimately fueled, at low radiative efficiency, by tidally 
driven accretion. 

Ground-based adaptive-optics imaging of near-IR H$_2$ {\it rovibrational} transitions in 3C 326 N will probe 
the spatial distribution of hot H$_2$ and further test the scenario of tidally induced accretion. The proposed 
H$_2$ Explorer mission \citep[H2EX,][]{bm07}, sensitive to H$_2$ emission over a large field of view, can 
establish the space density of the general MOHEG population and help determine its importance as a stage in 
galaxy evolution.

\section{Conclusions}

We have discovered extremely luminous molecular hydrogen pure-rotational emission lines from radio galaxy 
3C 326 N. The continuum emission is relatively weak, so that the H$_2$ emission contributes $17\%$ of the 
8-70 $\mu$m IR luminosity and is a major coolant. We measure a warm molecular gas mass of $1.1\times 10^9 ~M_\odot$ at 
temperatures of 125-1000 K within a 3.1 kpc radius of the galaxy center. The H$_2$ ortho/para ratios are consistent with
thermodynamic equilibrium. 

We consider 3 ways to heat the H$_2$ to the observed temperatures and produce the extreme H$_2$ emission 
line luminosities. The LINER AGN in 3C 326 N is most likely not luminous enough to power the observed H$_2$ emission via 
X-ray heating. Jet-shock heating could work in principle, but there does not appear to be enough kinetic energy 
dissipated inside the host galaxy by the jet. We favor the idea that accretion shocks from inflow induced by tidal 
interaction with 3C 326 S may power the remarkable H$_2$ emission spectrum.

Weak PAH emission features and thermal dust continuum in 3C 326 N and 3C 326 S indicate very little star formation 
activity $\sim 0.1-0.3 ~M_\odot$ yr$^{-1}$. The weak star formation in 3C 326 N and low accretion luminosity of the AGN 
are remarkable considering the large amount and high surface density of molecular gas. Similar H$_2$ emission seen in 
some ULIRGs is usually accompanied by vigorous starburst or AGN activity. The newly discovered molecular hydrogen 
emission galaxy (MOHEG) phenomenon may represent an important stage in galaxy evolution. Star-formation as well as accretion 
onto the central supermassive black hole may be suppressed by high temperatures, turbulence, or velocity shear in the large 
central mass of shocked molecular gas.

\acknowledgements

This work is based on observations made with the {\it Spitzer} Space Telescope, which is operated 
by the Jet Propulsion Laboratory (JPL), California Institute of Technology (Caltech) under NASA contract. Support for 
this research was provided by NASA through an award issued by JPL/Caltech. This research has made use of the 
NASA/IPAC Extragalactic Database (NED) which is operated by JPL/Caltech, under contract with NASA.
The IRAC 3-color image was processed using Montage, funded by NASA's Earth Science Technology Office, Computation Technologies 
Project, and maintained by the NASA/IPAC Infrared Science Archive (IRSA). We made use of data from the Sloan Digital Sky Survey 
funded by the Alfred P. Sloan Foundation, Participating Institutions, NSF, NASA, U.S. DOE, Japanese Monbukagakusho,  
Max Planck Society, and the Higher Education Funding Council for England. Thanks to K. Sheth for enlightening discussions on 
star formation in galaxies. Thanks also to T. Jarrett and R.-R. Chary for help with IRAC.


\clearpage

\begin{deluxetable}{lcccccccc}
\tabletypesize{\footnotesize}
\tablecaption{H$_2$ 0-0 S(J) Emission Lines}
\tablewidth{0pt}
\tablehead{
 & \colhead{H$_2$ S(0)} &  \colhead{ S(1)} &  \colhead{ S(2)} &  \colhead{ S(3)} &  \colhead{ S(4)} &  \colhead{ S(5)} &  \colhead{ S(6)} &  \colhead{ S(7)}}
\startdata
$\lambda$($\mu$m, rest) &  28.22      &  17.04      &  12.28      &  9.66       &  8.03       &  6.91       &  6.11         &  5.51         \\
 Flux\tablenotemark{a}  &  0.24(0.03) &  0.58(0.05) &  0.39(0.04) &  1.18(0.04) &  0.28(0.06) &  0.86(0.15) &  0.24(0.08)   &  0.28(0.08)   \\
 EW\tablenotemark{b}    &  1.9(0.6)   &  1.3(0.2)   &  0.41(0.08) &  0.80(0.08) &  0.14(0.03) &  0.34(0.10) &  0.074(0.025) &  0.066(0.022) \\
 Luminosity\tablenotemark{c} & 0.48   &  1.1        &  0.77       &  2.34       &  0.55       &  1.7        &  0.48         &  0.55         \\
 $N/g$\tablenotemark{d} &  9.2E18     &  2.8E17     &  3.8E16     &  6.9E15     &  1.3E15     &  4.4E14     &  1.5E14       &  2.7E13       \\  
\enddata
\tablenotetext{a}{Flux ($10^{-14}$ erg s$^{-1}$ cm$^{-2}$).}
\tablenotetext{b}{Equivalent Width ($\mu$m, rest).}
\tablenotetext{c}{~Luminosity ($10^{41}$ erg s$^{-1}$), at a luminosity distance of 407 Mpc.}
\tablenotetext{d}{Column Density (cm$^{-2}$) of upper level divided by its statistical weight, 
                  assuming a $3\farcs7\times 3\farcs7$ source.}
\end{deluxetable}

\clearpage

\begin{deluxetable}{lccccccccc}
\tabletypesize{\footnotesize}
\tablecaption{Forbidden Emission Lines}
\tablewidth{0pt}
\tablehead{
 \colhead{ } &\colhead{ [Fe II]} &  \colhead{  [S IV]} &  \colhead{ [Ne II]} &  \colhead{[S III] } &  \colhead{  [O IV]} }
\startdata
 $\lambda$($\mu$m, rest) &   5.34       &  10.51      & 12.81       &  18.71      &  25.89      \\
 Flux\tablenotemark{a}     & 0.20(0.08) &  0.08(0.04) &  0.40(0.04) &  0.14(0.07) &  0.25(0.03) \\
 Flux/[Ne II] &  0.5(0.2)  &  0.2(0.1)   &  1.0         &  0.4(0.2)  &  0.6(0.1) \\
\enddata
\tablenotetext{a}{Flux ($10^{-14}$ erg s$^{-1}$ cm$^{-2}$).}
\end{deluxetable}

\clearpage

\begin{deluxetable}{cccccccccc}
\tabletypesize{\footnotesize}
\tablecaption{Aromatic Emission Features (PAHs)}
\tablewidth{0pt}
\tablehead{
 \colhead{$\lambda(\mu$m)}   &\colhead{  6.7} &  \colhead{  7.7} &  \colhead{  10.7} &  \colhead{  11.3} &  \colhead{  13.6}&  \colhead{  15.9} &  \colhead{  17 } &  \colhead{  18.9 }}
\startdata
3C 326 N                    & 1.13(0.48)     &  0.53(0.26)      &  0.22(.09)        & 0.64(0.13)        &  0.98(0.18)      &  0.21(0.11)       & 1.15(0.27)       &  0.33(0.18) \\
3C 326 S                    & \nodata        &  1.61(0.42)      &  \nodata          &  1.19(0.20)       &  \nodata         &  \nodata          &  \nodata         & \nodata     \\
\enddata
\tablenotetext{a}{Flux ($10^{-14}$ erg s$^{-1}$ cm$^{-2}$).}
\end{deluxetable}

\clearpage

\begin{deluxetable}{ccccccc}
\tabletypesize{\footnotesize}
\tablecaption{MIPS 24, 70, and 160 $\mu$m Photometry}
\tablewidth{0pt}
\tablehead{  \colhead{ }  &  \colhead{$F_\nu$(24 $\mu$m)\tablenotemark{a}} & \colhead{$F_\nu$(70 $\mu$m)\tablenotemark{b}}
           & \colhead{$F_\nu$(160 $\mu$m)\tablenotemark{c}} &
             \colhead{L(24 $\mu$m)\tablenotemark{d}} & \colhead{L(70 $\mu$m)} & \colhead{L(70)/L(24)}}
\startdata
3C 326 N  & 0.52$\pm$0.07 & 6.12$\pm$0.65 & $<34$ & 1.3$\pm$0.2  & 5.2$\pm$0.6 &  4.0$\pm$0.8 \\
3C 326 S  & 1.05$\pm$0.12 & 4.33$\pm$0.60 & $<34$ & 2.6$\pm$0.3  & 3.7$\pm$0.5 &  1.4$\pm$0.3 \\
bridge\tablenotemark{e} & \nodata  & 4.40$\pm$0.60 &   \nodata  &   \nodata   &  3.8$\pm$0.5 & \nodata \\
\enddata
\tablenotetext{a}{Flux density (mJy) in the 24 $\mu$m band, measured in a $4\farcs9$ radius aperture and multiplied by
                  an aperture correction of 2.0.} 
\tablenotetext{b}{Flux density (mJy) in the 70 $\mu$m band, measured in a $9\farcs84$ radius aperture and multiplied by
                  an aperture correction of 1.5.}
\tablenotetext{c}{The $3\sigma$ upper limit to the 160 $\mu$m flux density (mJy), measured in a $20\arcsec$ radius 
                  aperture.}
\tablenotetext{d}{~Luminosity $\nu L_\nu$($10^{42}$ erg s$^{-1}$), at a luminosity distance of 407 Mpc.}  
\tablenotetext{e}{This source (only in the MIPS 70 $\mu$m image) is found at roughly the location
                  where the tidal bridge appears to connect to 3C 326 N (Fig. 5).}
\end{deluxetable}

\clearpage

\begin{deluxetable}{ccccc}
\tabletypesize{\footnotesize}
\tablecaption{H$_2$ Model Parameters}
\tablewidth{0pt}
\tablehead{\colhead{T(K)} & \colhead{$M$(H$_2$)\tablenotemark{a}} & \colhead{Ortho/Para} & \colhead{$n_\mathrm{crit}$\tablenotemark{b}} & \colhead{$f_\mathrm{max}$\tablenotemark{c}}}
\startdata
125  & 1.1E9 & 2.1 & 5.3E2 & 1.5E-3 \\
400  & 1.1E7 & 3.0 & 1.1E4 & 7.0E-7 \\
1000 & 6.4E5 & 3.0 & 4.4E5 & 1.0E-9 \\
\enddata

\tablenotetext{a}{H$_2$ mass ($M_\odot$).} 
\tablenotetext{b}{Critical density (cm$^{-3}$).}  
\tablenotetext{c}{Maximum filling factor under LTE conditions.}                                                                    
\end{deluxetable}

\clearpage

\begin{deluxetable}{lccccc}
\tabletypesize{\footnotesize}
\tablecaption{Strong H$_2$ Emitters}
\tablewidth{0pt}
\tablehead{  \colhead{ }  & \colhead{M(cold H$_2$)\tablenotemark{a}} & \colhead{M(warm H$_2$)\tablenotemark{b}} & 
             \colhead{$L(H_2)$\tablenotemark{c}} & \colhead{$L(H_2)/L(IR)$\tablenotemark{d}} & \colhead{Refs.}}
\startdata
Zw 3146 BCG       & $\sim 1\times 10^{11}$ & $\sim 1\times 10^{10}$ & $2.1\times 10^{43}$ & 0.014  & 1   \\
NGC 6240          &    $1.1\times 10^{10}$ & $1.6\times 10^{9}$ & $4.6\times 10^{42}$ & 0.003  & 2,3 \\
3C 326 N          &    \nodata             & $1.1\times 10^{9}$ & $8.0\times 10^{41}$ & 0.04-0.1 &     \\
Antennae          &    $9.6\times 10^{9}$  & $4.9\times 10^{8}$ & $1.7\times 10^{41}$ & 0.0009 & 4,5,6 \\
Stephan's Quintet &    \nodata             & $>3.4\times 10^{7}$  & $>8.4\times 10^{40}$ & 0.34   & 7   \\
\enddata
\tablenotetext{a}{Cold H$_2$ mass derived from CO measurements.}
\tablenotetext{b}{Warm H$_2$ mass measured from Spitzer or ISO spectra.}
\tablenotetext{c}{H$_2$ Luminosity (erg s$^{-1}$).}  
\tablenotetext{d}{The IR luminosity is integrated over 8-1000 $\mu$m.}  
\tablerefs{(1) Egami et al. 2006; (2) Lutz et al. 2003; (3) Armus et al. 2006; 
                              (4) Haas et al. 2005; (5) Young et al. 1995; (6) Klaas et al. 1997; (7) Appleton et al. 2006.}
\end{deluxetable}


\clearpage

\begin{figure}[ht]
   \plotone{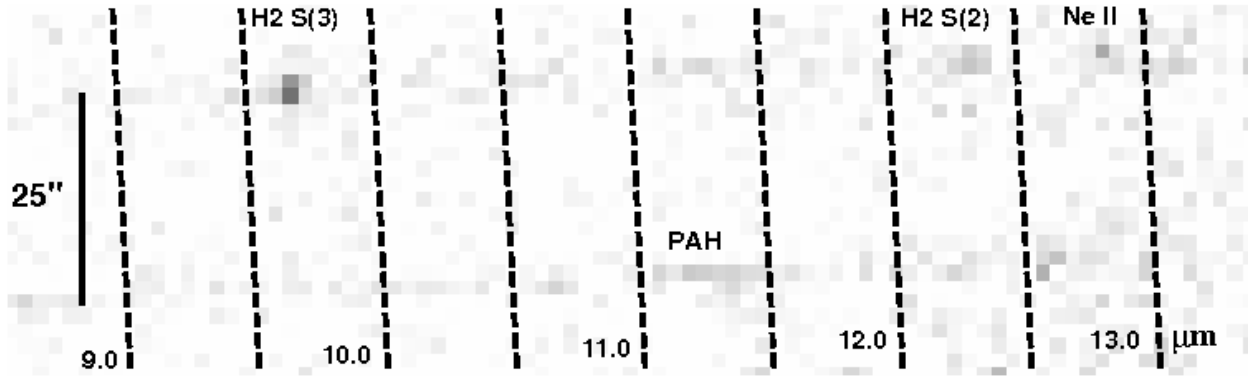}
   \figcaption{{\it Spitzer} IRS SL1 8.5-13.4 $\mu$m (rest) 2D spectra of 3C 326 N ({\it top}) and 3C 326 S ({\it bottom}). 
               The bright H$_2$ S(3) 9.66 $\mu$m emission line in 3C 326 N is spatially and spectrally unresolved. The H$_2$ 
               S(2) and Ne {\sc ii} lines are also seen. The southern galaxy was serendipitously placed precisely on the SL 
               slit (PA$=178\arcdeg$). Its 11.3 $\mu$m PAH emission feature is indicated.}
\end{figure}

\clearpage

\begin{figure}[ht]
   \plotone{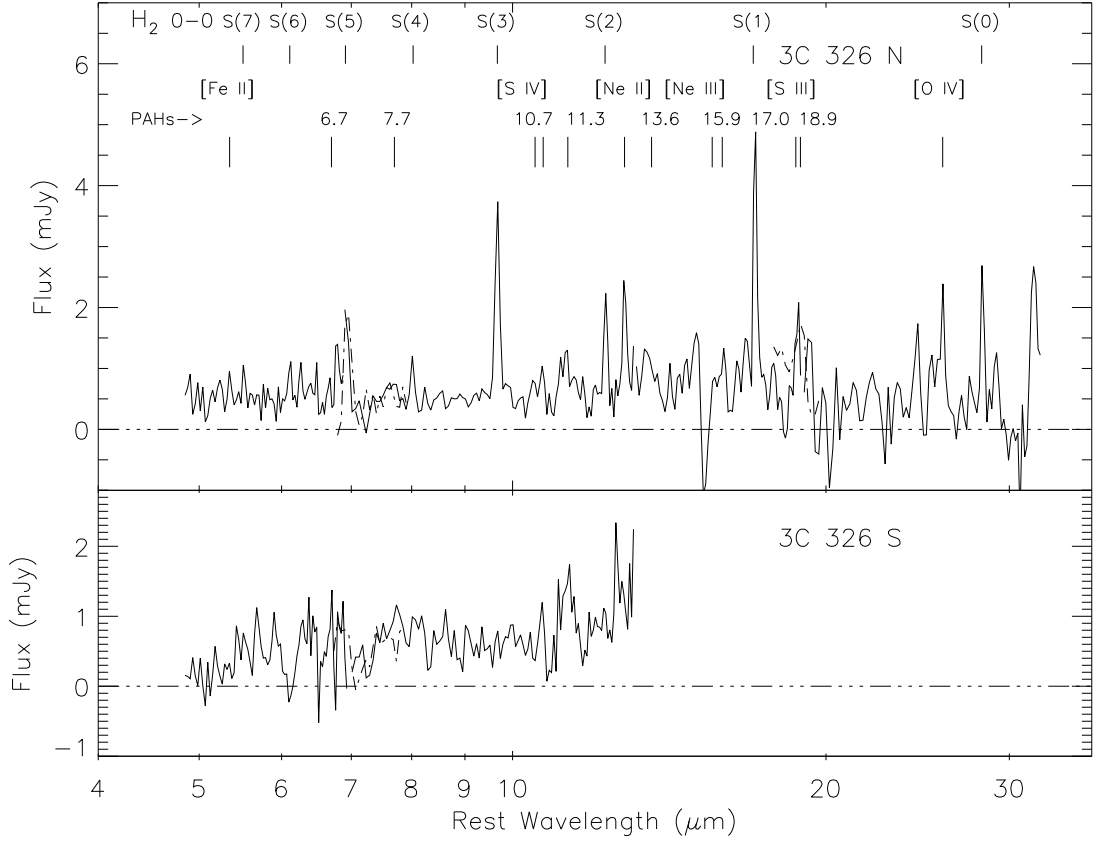}
   \figcaption{{\it Top}: {\it Spitzer} IRS spectrum of 3C 326 N. {\it Dashed lines}--SL3 and LL3 bonus orders. Extremely
               strong H$_2$ emission lines of high equivalent width dominate the spectrum. Weak PAH
               features indicate a very low level of star formation in spite of the large amount of H$_2$. Forbidden
               ionic emission line ratios are consistent with a weak LINER AGN. {\it Bottom}: PAH emission is detected 
               in 3C 326 S, but no H$_2$ or ionic forbidden lines.}
\end{figure}

\clearpage

\begin{figure}[ht]
   \plotone{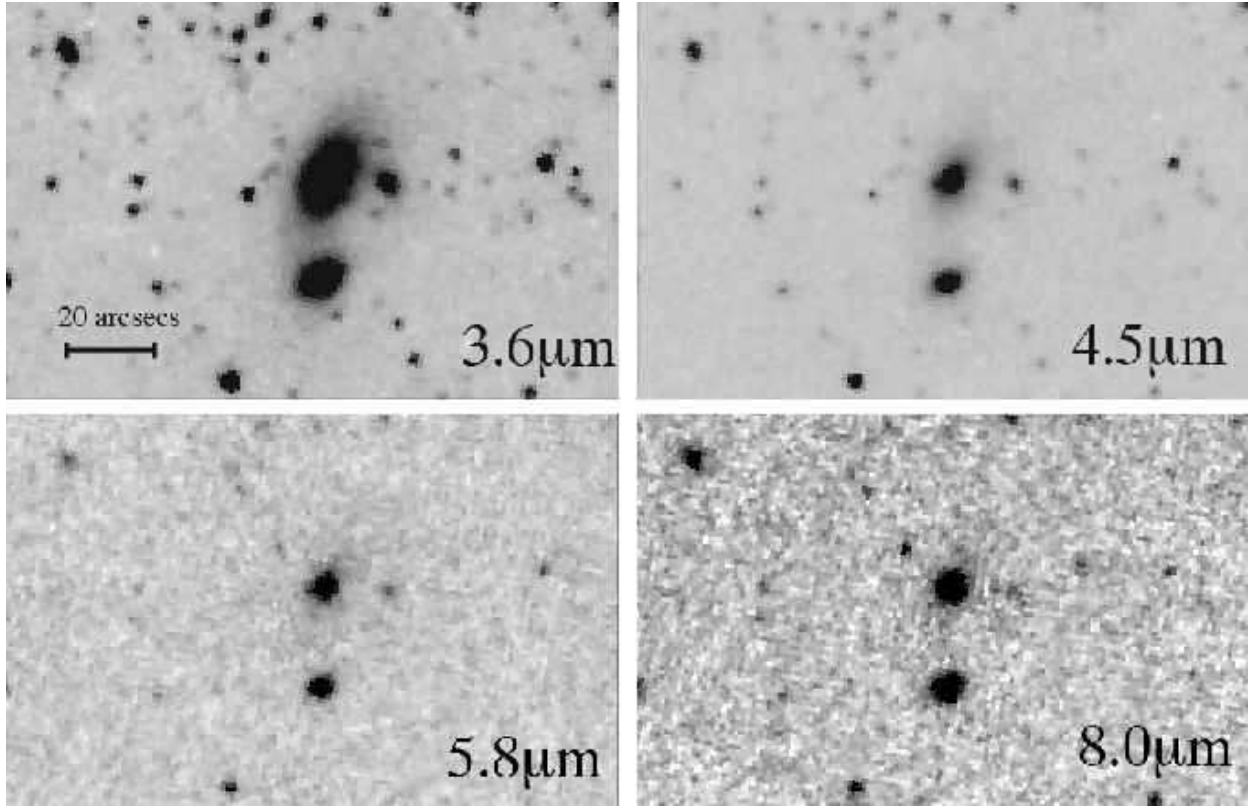}
   \figcaption{{\it Spitzer} IRAC 3.6, 4.5, 5.8, and 8.0 $\mu$m images of 3C 326. North 
                  is at top, East to the left. The first two bands are dominated by stellar emission,
                  while the second two have contributions from dust and H$_2$. The galaxies 3C 326 N and 3C 326 S 
                  are the brightest two sources in these images. A third source at PA $=268\arcdeg$
                  and a fourth $4\farcs3$ away from 3C 326 N at PA $=111\arcdeg$ (seen protruding from 
                  the left of the galaxy core in the 4.5 $\mu$m image) appear to be unrelated galaxies seen in 
                  projection (see \S 3.3). Note the apparent tidal bridge between  3C 326 N and 
                  S and a possible tidal tail to the NW in the 3.6 and 4.5 $\mu$m images, evidence of an ongoing 
                  galaxy collision. The galaxy cores are barely resolved at 5.8 $\mu$m, and 3C 326 S is nearly 
                  as bright as 3C 326 N at 8.0 $\mu$m.}
\end{figure}

\clearpage

\begin{figure}[ht]
   \plotone{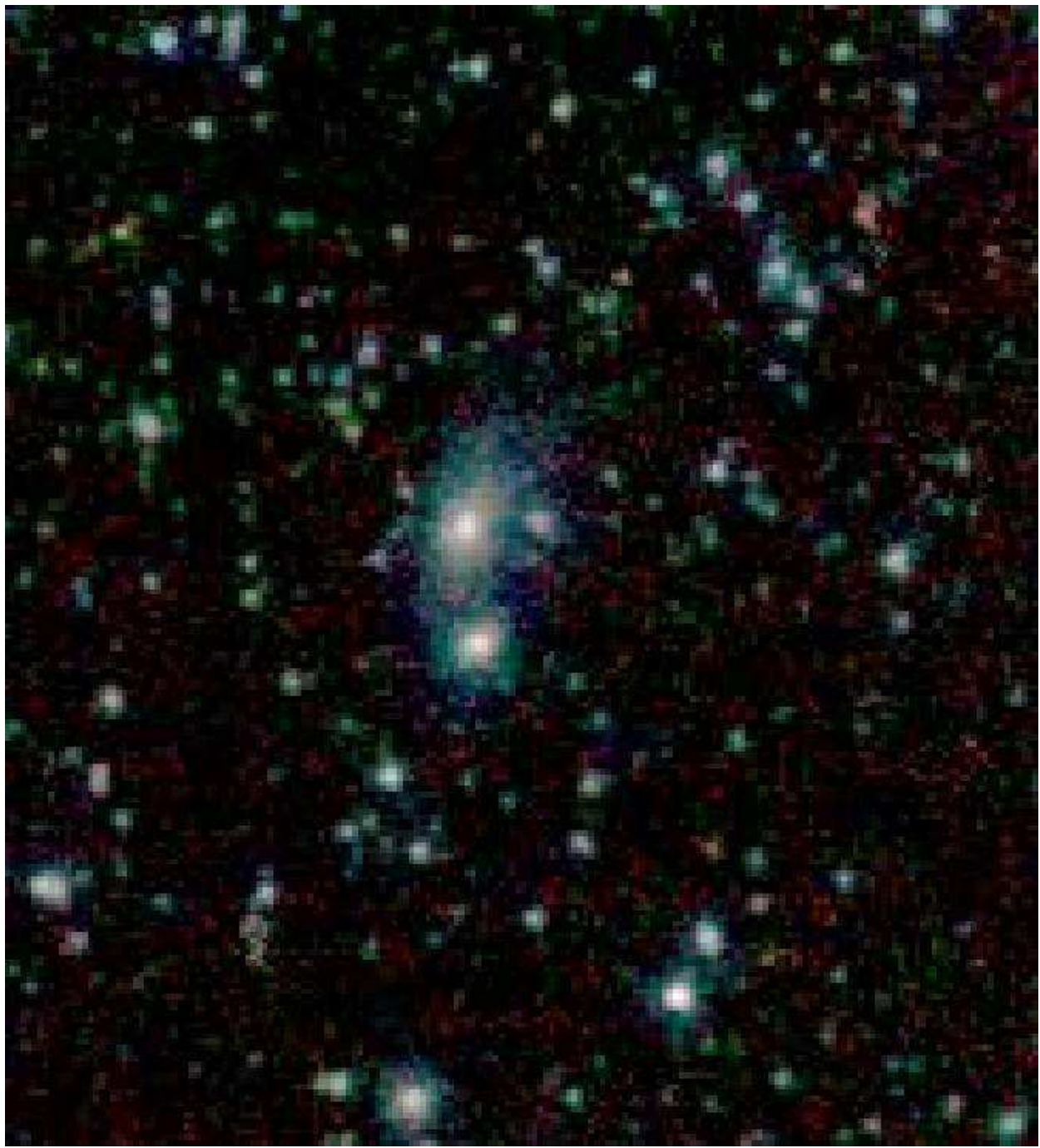}
   \figcaption{{\it Spitzer} IRAC 3-color image of 3C 326. Blue $=$ 3.6 $\mu$m, green=4.5 $\mu$m, red=5.8 $\mu$m. 
               Foreground stars and starlight-dominated galaxies appear blue-green, while dust or H$_2$ emission 
               appears red. The brightest galaxy and AGN host 3C 326 N has an irregular morphology because of 
               its interaction with 3C 326 S. Note the possible tidal tail extending to the NW.}
\end{figure}

\clearpage

\begin{figure}[ht]
   \plotone{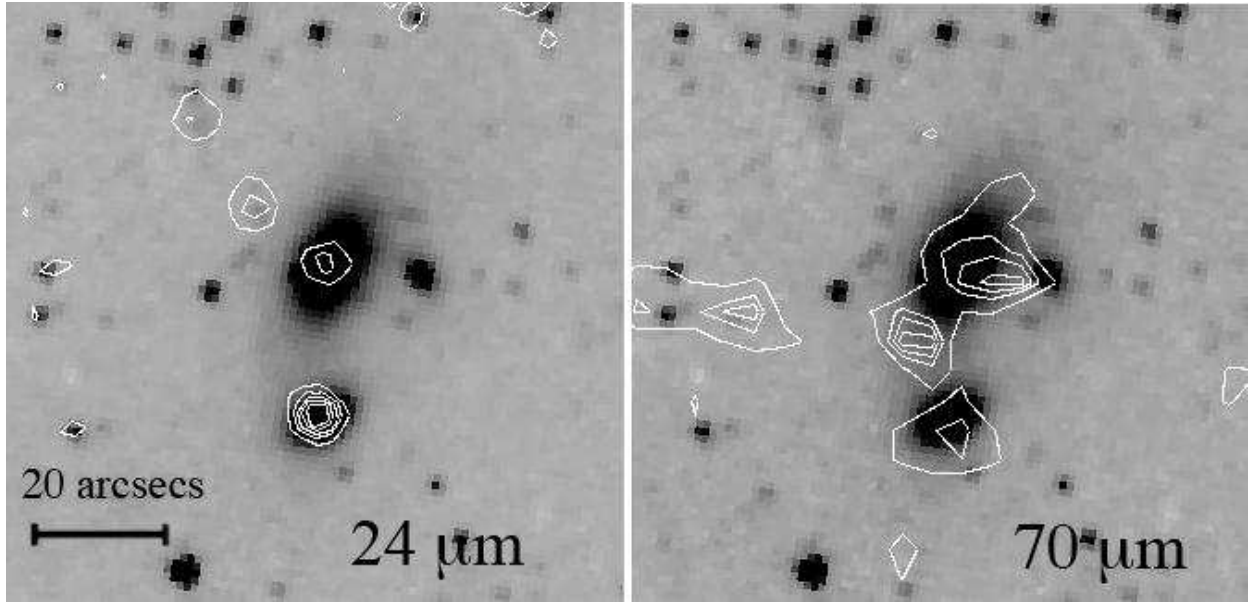}
   \figcaption{{\it Left}: {\it Spitzer} MIPS 24 $\mu$m flux contours overlaid on the IRAC 3.6 $\mu$m image of 3C 326. North 
                  is at top, East to the left. The contour levels are 0.3, 0.4, 0.5, and 0.6 MJy/sr
               {\it Right}: MIPS 70 $\mu$m contours overlaid on the 3.6 $\mu$m image.  The contour levels are 0.45, 0.65, 0.75
                  and 0.85 MJy/sr. 3C 326 S is brighter at 24 $\mu$m, while 3C 326 N is brighter at 70 $\mu$m. A third 70 
                  $\mu$m source appears between the two galaxies, but not in the IRAC image. It may be cool dust emission 
                  associated with the tidal bridge, or an unrelated background source.}
\end{figure}

\clearpage

\begin{figure}[ht]
   \plotone{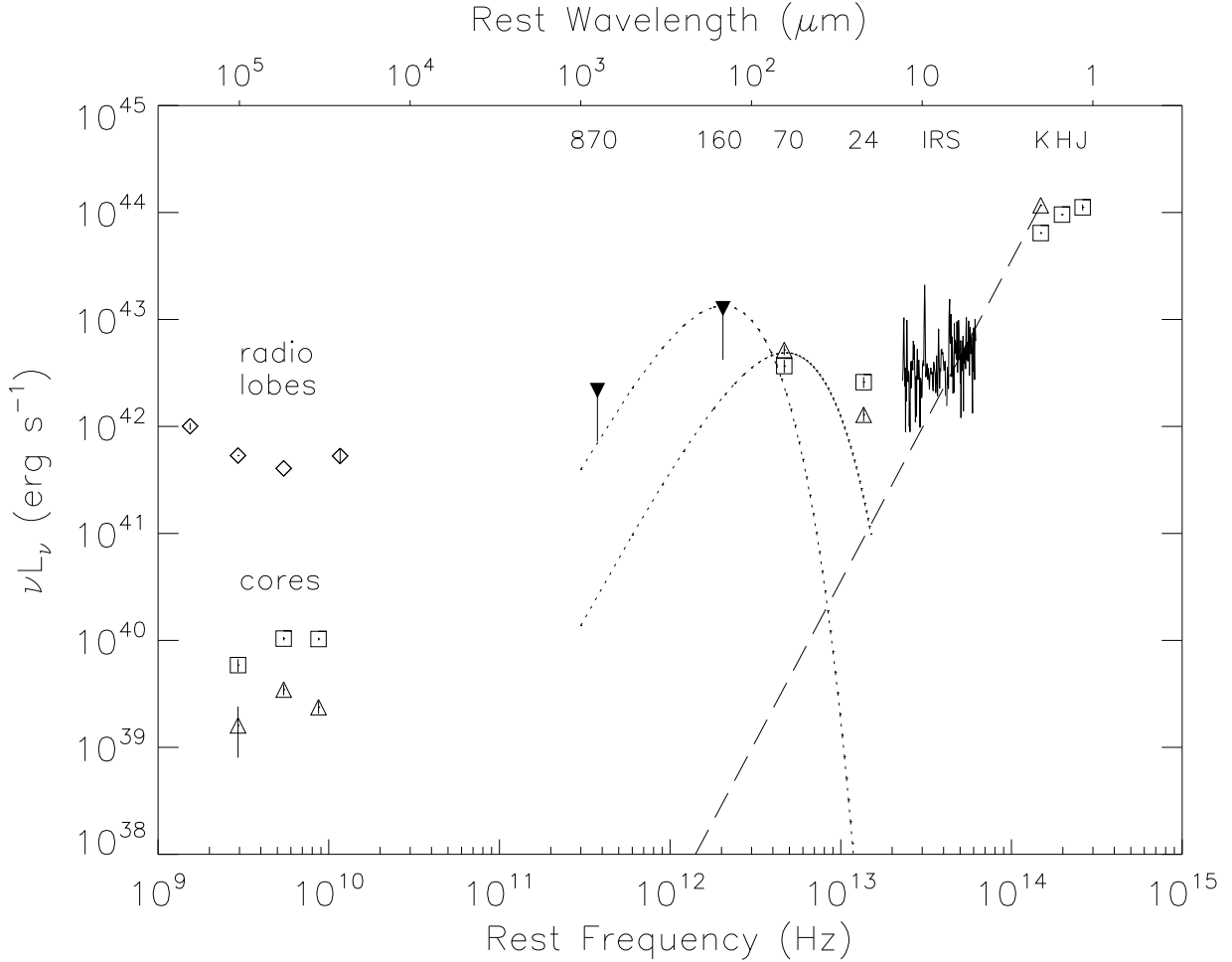}
   \figcaption{Spectral energy distributions of 3C 326 N ({\it open triangles and solid line}), 3C 326 S ({\it squares}), and 3C 326 
               radio lobes ({\it diamonds}). MIPS 24 and 70 $\mu$m points indicate warm and cold dust components. MIPS 160 $\mu$m 
               and ground-based 870 $\mu$m \citep{qay03} $3\sigma$ upper limits ({\it inverted triangles}) apply to both galaxies 
               individually. J, H, and K-band near-IR photometry are by \cite{ll84}. The radio core fluxes and the K-band photometry 
               of 3C 326 N are from \cite{rsm90}. Radio lobe data are from \cite{ws78} and the NASA/IPAC Extragalactic Database (NED). 
               The dashed line represents the Rayleigh-Jeans tail of starlight connecting the K-band photometry to the IRS spectrum of 
               3C 326 N. The dotted lines are 25 and 58 K blackbodies, representing upper and lower limits to
               the 70-1000 $\mu$m emission.}
\end{figure}

\clearpage

\begin{figure}[ht]
   \plotone{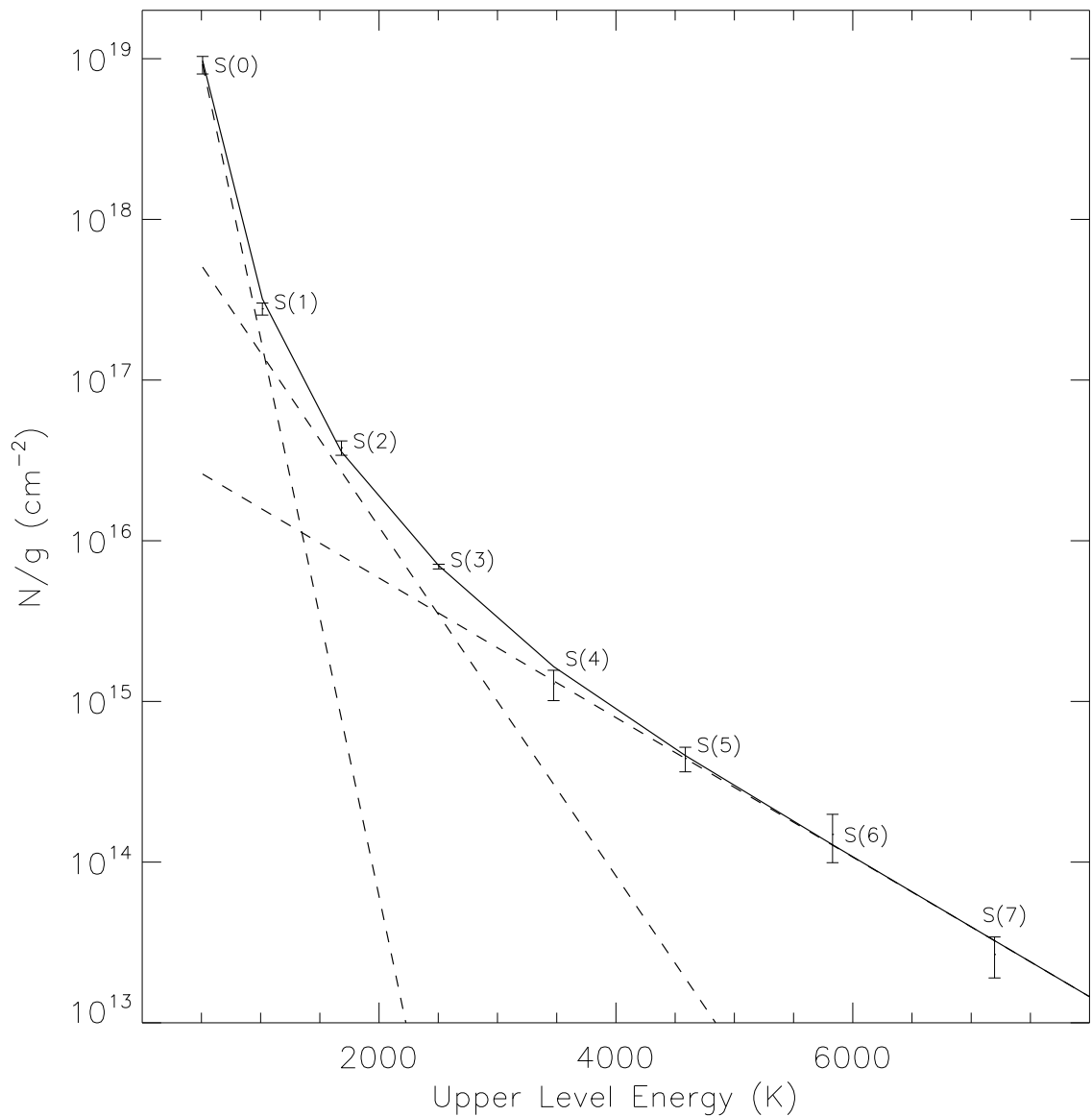}
   \figcaption{H$_2$ excitation diagram for 3C 326 N. Upper-level column densities normalized 
               by their statistical weights are plotted as a function of upper-level energy ($E_\mathrm{u}/k_\mathrm{B}$) 
               for each pure-rotational transition. The best-fitting model ({\it solid line}) has 3 temperature components
               ({\it dashed lines}): 125, 400, and 1000 K. The ortho/para ratios are set at the respective thermal 
               equilibrium values (Table 5). Note that the y-intercept for each model component gives the average
               H$_2$ ground-state column density integrated over the SL slit.} 
 \end{figure}

\end{document}